# Correlative and *in situ* microscopy investigation of phase transformation, crystal growth and degradation of antimony sulfide thin films

Mingjian Wu[1]*, Maïssa K. S. Barr[2], Vanessa M. Koch[2], Martin Dierner[1], Tobias Dierke[3], Penghan Lu[4], Johannes Will[1], Rafal Dunin-Borkowski[4], Janina Maultzsch[3], Julien Bachmann[2] and Erdmann Spiecker[1]*

1 Institute of Micro- and Nanostructure Research & Center for Nanoanalysis and Electron Microscopy (CENEM) Friedrich-Alexander-Universität Erlangen-Nürnberg, IZNF, Cauerstraße 3, 91058 Erlangen, Germany

2 Chemistry of Thin Film Materials, Department Chemistry and Pharmacy, Friedrich-Alexander-Universität Erlangen-Nürnberg, IZNF, Cauerstraße 3, 91058 Erlangen, Germany

3 Chair of Experimental Physics, Department of Physics, Friedrich-Alexander-Universität Erlangen-Nürnberg, Staudtstraße 7, 91058 Erlangen, Germany

4 Ernst Ruska-Centre for Microscopy and Spectroscopy with Electrons, Research Centre Jülich, 52425 Jülich, Germany

Corresponding Emails: mingjian.wu@fau.de; erdmann.spiecker@fau.de;

**Abstract**

Antimony sulfide ($Sb_2S_3$), a compound of earth-abundant elements with highly anisotropic, quasi-layered crystal structure, triggered growing interest as a solar absorber in photovoltaics and as a phase change material in memory devices, yet challenges remain in achieving high-quality thin films with controlled nucleation and growth for optimal performance. Here, we investigate the phase transformation, crystal structure and properties, growth and degradation of atomic layer deposited $Sb_2S_3$ thin films using *in situ* TEM and correlative *ex situ* analysis. The as-deposited amorphous films crystallized at 243°C, forming grains with an [100] out-of-plane texture and developed into tens to hundreds of micrometer long, leaf shaped grains. Introducing an ultra-thin ZnS interfacial layer increased nucleation density, resulted in few micrometer sized, more uniform grains, while retaining the overall [100] texture. *In situ* observations and subsequent crystal orientation analysis with cutting-edge 4D-STEM and EBSD revealed that the grains grew faster along the [010] ribbon direction and that the bare films underwent early-stage degradation, forming holes in amorphous regions during annealing. The ZnS interlayer mitigated degradation, stabilizing the films and improving their uniformity. These findings offer valuable insights for optimizing $Sb_2S_3$ thin films for applications both as solar cell materials and phase change materials.

1. **Introduction**

Antimony sulfide ($Sb_2S_3$) and related antimony dichalcogenides have attracted considerable interest for optoelectronic and photovoltaic applications due to their favorable semiconductor properties, including a direct bandgap, high absorption coefficient, and chemical stability.[1-10] These characteristics make $Sb_2S_3$ a promising material for solar cells, photodetectors, and thin-film transistors. The ability to process $Sb_2S_3$ into high-quality thin films is crucial for optimizing device performance, yet challenges remain in controlling its crystallization and film morphology.

Beyond optoelectronics, $Sb_2S_3$ is increasingly explored as a phase change material (PCM) for non-volatile memory and thermal switching devices.[11-17] Its rapid and reversible phase transformations between amorphous and crystalline phases, coupled with a significant contrast in electrical and optical properties, makes it a potential alternative to the conventional PCMs like the Ge-Sb-Te family.[18] Despite its promise, the fundamental understanding of the crystallization mechanisms of $Sb_2S_3$, including the role of structural anisotropy and interface effects, remains limited. Detailed insight into the mechanisms of nucleation, grain growth, and phase stability are critical for advancing its practical applications in both fields of optoelectronics and PCM.

A key aspect influencing device performance is the orientation of $Sb_2S_3$ crystals. It crystallizes in a highly anisotropic structure, forming covalently bonded quasi-one-dimensional ribbons. Charge transport is significantly more facile along the ribbon direction than in the perpendicular directions.[6-8, 10, 14] Therefore, controlling the crystal orientation in fundamentally anisotropic devices such as solar cells might provide an important avenue for performance optimization. If the ribbon orientation can be engineered to coincide with the desired carrier collection direction, losses can be minimized, perhaps even including losses caused by defects. Furthermore, maintaining precise control over the film morphology is equally important, as the spontaneous formation of elongated, needle-like structures along the ribbon direction, frequently observed in both microscopic and macroscopic crystals, can hinder the performance of semiconductor junction-based devices.

Atomic layer deposition (ALD) provides precise control over film thickness and composition, enabling the growth of highly uniform $Sb_2S_3$ thin films on arbitrary surfaces.[19, 20] However, as-deposited $Sb_2S_3$ films are typically amorphous and require thermal annealing to crystallize into the desired crystalline phase. The crystallization process is influenced by factors such as substrate properties, temperature, and the presence of interfacial layers, all of which affect the resulting texture, crystal size, and overall film quality. Despite significant progress, the underlying mechanisms governing nucleation, growth, and degradation during this transformation remain insufficiently understood.

This study combines *in situ* transmission electron microscopy (TEM) with correlative microscopy and spectroscopy to investigate the crystallization, grain growth, and degradation of ALD-deposited $Sb_2S_3$ thin films. We focus on elucidating how structural anisotropy, substrate curvature, and interfacial layers (specifically a ZnS adhesion layer) influence these processes. Using *in situ* TEM, we observe the amorphous-to-crystalline phase transition, track growth kinetics, and identify early-stage degradation phenomena. Complementary techniques, including 4D-STEM and Electron Backscatter Diffraction (EBSD), provide insights into the grain texture and orientation across different substrates. Furthermore, polarized light microscopy and polarization-dependent Raman spectroscopy are applied correlatively with structural information derived from TEM/SEM to elucidate the orientation-dependence of the vibrational modes.

## 2. Results and discussions

***Phase transformation of $Sb_2S_3$.*** Figure 1(a-d) illustrates the phase evolution of ALD-deposited antimony sulfide thin films, as observed using bright-field TEM (BF-TEM) and selected area electron diffraction (SAED). The as-deposited films are amorphous (a-$Sb_2S_3$), confirmed by the granular features in TEM imaging and the absence of sharp diffraction spots/rings in SAED (Fig. 1a). The presence of an ultrathin ZnS interfacial layer does not significantly alter the morphology or diffraction pattern of the as-deposited films. Its presence is, however, confirmed through the Zn signals in EDX spectra (Supplementary Fig. S1), which is absent in the thin films without the ZnS layer.

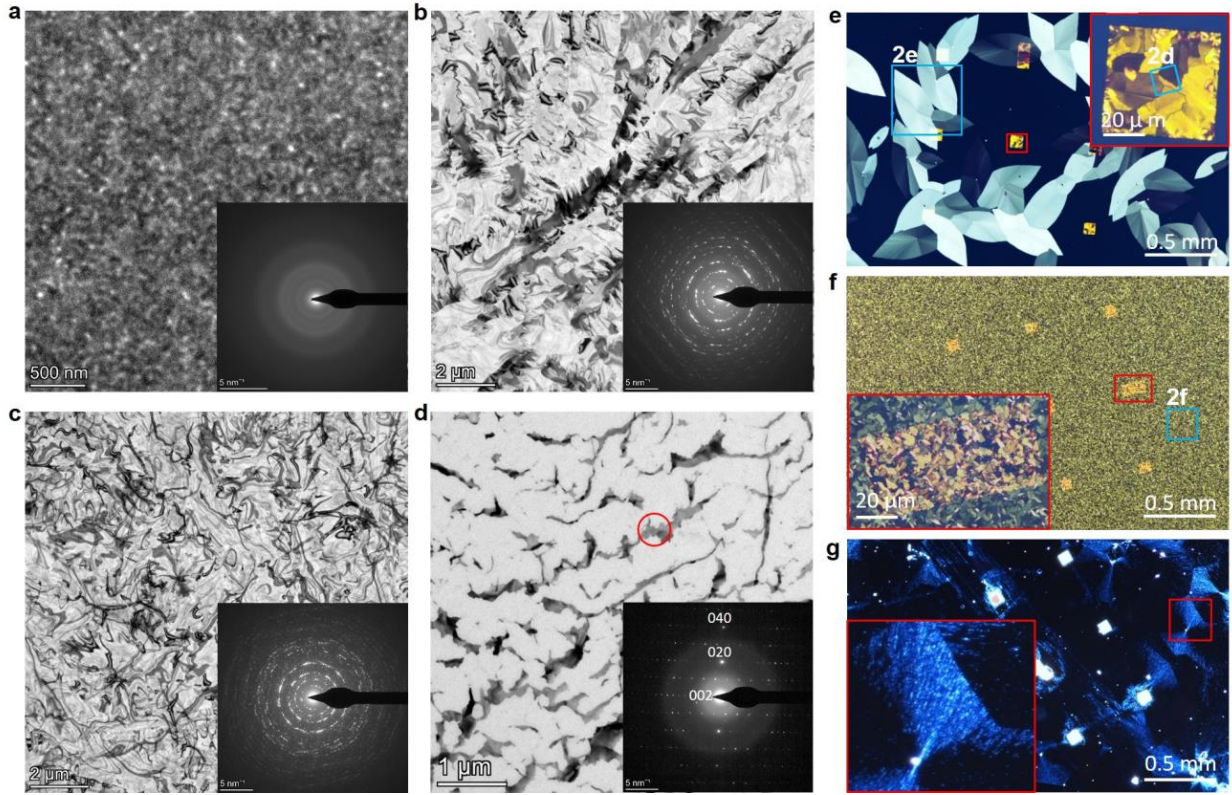

*Figure 1. (a-d) bright field (BF-)TEM images and selected area electron diffraction (SAED) patterns as insets of the antimony sulfide thin films at different deposition and annealing conditions. (a) The as-deposited a-$Sb_2S_3$ thin film. (b) The film in (a) after annealing at 250°C for 10 min in TEM. (c) the ZnS/a-$Sb_2S_3$ film after annealing at 250°C for 10 min in TEM. (d) The a-$Sb_2S_3$ thin film in the same batch as (a) after annealing at 350°C for 10 min in TEM. SA aperture for (a-c) included large ~13 μm diameter of the thin films, and only ~460 nm diameter in (d) as marked by the red circled region. (e) Polarized light microscopy (PLM) of the sample shown in (b) supported on Si/$SiO_x$ TEM grid. The yellowish rectangles/squares (one magnified as inset) are the 30nm $SiO_x$ membrane windows for TEM observations. (f) PLM of the sample shown in (c) with one membrane window magnified as inset. (g) dark-field optical micrograph of the sample shown in (d) revealing the traces of evaporated $Sb_2S_3$. Blue boxed regions in (e) and (f) are further analyzed and shown in Fig. 2.*

After annealing above 243°C, the amorphous thin films transform to crystalline phase, both with and without the ZnS interfacial layer. The diffraction contrast TEM images reveal crystal phase of $Sb_2S_3$ with strong bending contours (Fig. 1b-d, Fig. 2d and Fig. 3, further discussed later). This enables an easy differentiation of the crystal phase from the as-deposited amorphous phase, due to the strong diffraction contrast, either using BF-TEM, or the complementary BF-, annual bright field (ABF-) and annual dark field (ADF-) STEM imaging modalities, e.g., cf. supplementary Fig. S2 and videos. Whereas the high angle annual dark-field (HAADF-) STEM images suppresses the diffraction contrast and best reveal the crystal grain boundaries.

The amorphous to crystalline phase transformation temperature of $Sb_2S_3$ was carefully and consistently confirmed in five *in situ* annealing experiments (four experiments using a-$Sb_2S_3$ and one using ZnS/a-$Sb_2S_3$). In a first exploratory experiment, the temperature was increased by 50°C in each step and held for at least three minutes to observe possible phase transformation. Here the thin film crystallized after heating from 200 to 250°C and holding at 250°C. In a subsequent experiment, the temperature was first ramped to 200°C and then increased by 5°C per step. Here, the phase transformation could be narrowed down to take place between 240 and 245°C. The corresponding video documenting the phase transformation can be found in supplementary video 1, which is further analyzed in Fig. 5 and discussed later below. Consequently, we carried out an additional experiment, where the temperature was slowly ramped to 242°C and held there for 30 min. During this period, no indication of phase transformation was observed. Upon raising the temperature to 243°C, crystalline grains evolve within two minutes. Repeating this protocol for a ZnS/a-$Sb_2S_3$ thin film yields the same phase transformation temperature of 243°C for a-$Sb_2S_3$ to $Sb_2S_3$ in vacuum. Thus, our systematic *in situ* investigations reveal that the phase transformation takes place reproducibly and independent of the interlayer at a defined temperature of 243°C. The identified phase transformation temperature is in agreement with the range reported in literature [12-14] but much more precise.

***Role of the ZnS interlayer on crystal size.*** As described above, the ZnS interlayer plays a significant role in controlling the grain size. Without interlayer, grain sizes of up to 10 µm in length and few micrometers in lateral width are observed on the thin $SiO_x$ windows for TEM observations (Fig. 1b and inset in 1e). The $Sb_2S_3$ grains reach even hundreds of microns on flat surfaces of $Si/SiO_x$, which was not accessible by TEM, as revealed by polarized light microscopy (PLM, Fig. 1e). This difference in size, dependent on the underlying surface curvature (buckled freestanding $SiO_x$ membrane vs. flat native SiOx on Si), points to the influence of substrate topography on nucleation density. In contrast, the ZnS interlayer promotes a higher nucleation density, yielding smaller, more uniform grains of around 2-3 µm (Fig. 1c and f), which is independent of the underlying substrate topology. The results suggest that ZnS provides heterogeneous nucleation sites, dictating grain size and minimizing the significant variability observed in ZnS-free samples. Thus, the interlayer also leads to a much higher degree of in-plane isotropy due to the finer grain structure as compared to the micron sized leaf structure without ZnS.

***Evaporation of the $Sb_2S_3$ thin films.*** Upon further heating a $Sb_2S_3$ sample to 350°C, evaporation of the crystalline $Sb_2S_3$ film was observed, with grain boundaries initiating the degradation process (cf. supplementary video 2). The BF-TEM image and the corresponding indexed SAED pattern in Fig. 1d and as insets, respectively, highlight the preferred loss of materials along the [010] direction. This is also indicated by the aligned traces of residual film fragments in Fig. 1g. This suggests a potential path for sulfur volatilization at elevated temperatures in vacuum, consistent with previous findings on thermal instability in similar sulfide compounds.[14] In regions protected from vacuum exposure, such as the periphery of the TEM grid, the $Sb_2S_3$ film remained stable, illustrating the influence of environmental conditions on its stability.

***Texture of the $Sb_2S_3$ thin films.*** $Sb_2S_3$ is a biaxial, highly anisotropic material which crystallizes into an orthorhombic structure belonging to the Pnma (#62) space group with lattice parameter a = 1.131 nm, b = 0.386 nm and c = 1.123 nm [21], as shown in Fig. 2a. The structure can be viewed as $[Sb_4S_6]_n$ ribbons extended with strongest (shortest) covalent bonds along the [010] direction, and with weaker (longer) interactions along [100]. The weakest (longest) covalent bonds along [100] form a quasi-2D layered structure. This highly anisotropic nature of the crystal structure facilitates an anisotropic growth of the crystals. We note that in other studies, $Sb_2S_3$ is assigned to Pbnm (#62) space group where the axes are permutated thus the indices are differently referred.

The Sb$_2$S$_3$ crystals grown on the SiO$_x$ membrane windows mostly exhibit an elongated shape, which is best revealed by the HAADF-STEM images (cf. Fig. S2) via their crystal grain boundaries. Strong bending contours are also visible, which indicate a high stress after the thin film crystallization. Furthermore, periodic folds, i.e., the short lines perpendicular to the grain boundary seemingly "carving" into the crystals, along the long grain boundaries can be very often seen (cf. Fig. 1b, Fig. 2b, Fig. 5b and Fig. S2). Although the bending contours complicate TEM analysis of grains, they provide valuable information about the character of the bending and folds in the thin films, given that the images were acquired under controlled diffraction condition, which require defined sample tilt conditions, in either BF-TEM or ADF-STEM modality. Figure 2b highlights a long grain with periodic folds along the grain boundary. The SAED pattern taken from the red-circled region can be easily indexed to very close to the [100] zone axis. The diffraction vector and corresponding image confirm the long axis of the crystal was grown along the [0k0] i.e., the ribbon direction as indicated by the green arrows. Along the long axis at the center of this crystal, a more homogeneous contrast is revealed compared to that close to the folds, indicating the crystals are flat in the middle and are bent/inclined closer to the folds.

The contrast at the periodic folds at the edge along the grain boundary is quite different. One of the fold marked by the blue box in Fig. 2b is magnified and shown as inset on top. We further examine the region across the fold using HRTEM lattice imaging (in Fig. 2c). The {002} lattice with 0.56 nm spacing is clearly revealed, whereas the {002} lattice could only be revealed on the top-right side via the fast Fourier transform (FFT, inset) pattern, whereas the lower-left region only preserved the {00*l*} system row of lattice as seen in the FFT due to bending and tilt out of the zone axis. This indicates that the ribbons in the film across the folds bend and tilt inward or outward of the viewing direction, i.e., ***along*** the ribbon direction. This characteristic bending is likely a consequence of the crystal anisotropy, the growth kinetics and stress relaxation mechanisms.

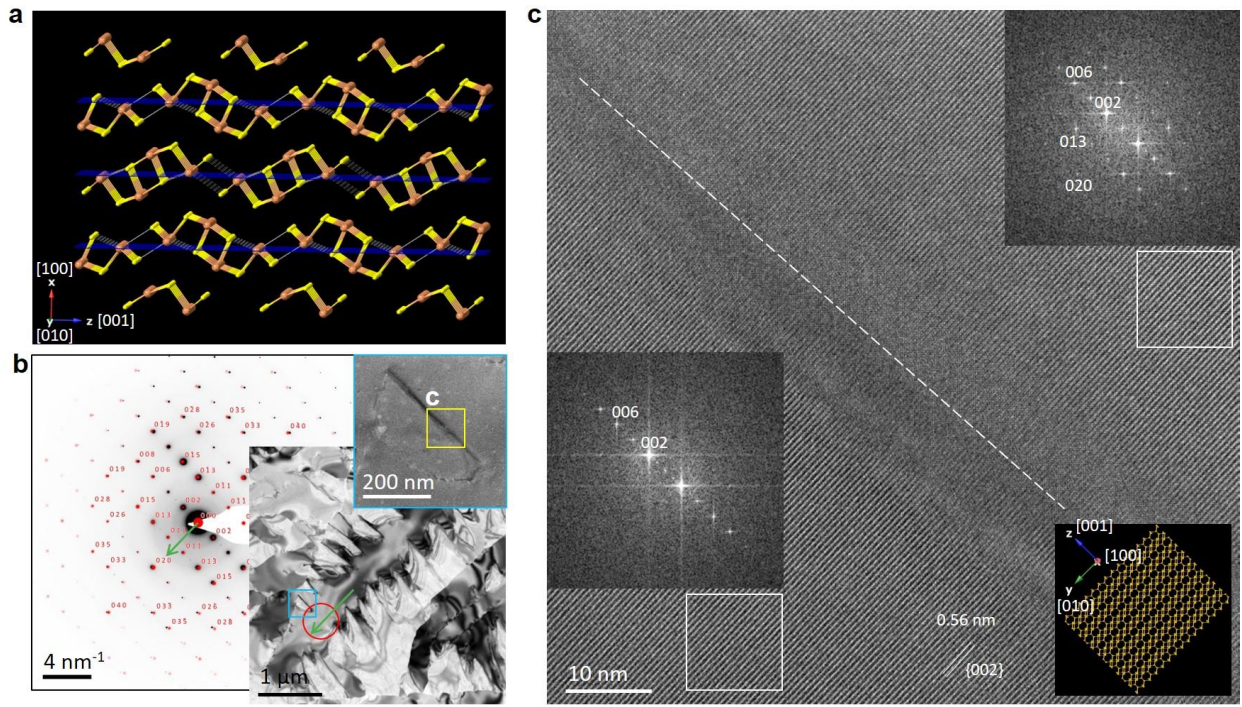

*Figure 2 (a) crystal structure of Sb$_2$S$_3$, (b) BF-TEM image highlights a long Sb$_2$S$_3$ crystal and the SAED from the red circled region with simulated pattern overlay, confirming [100] Sb$_2$S$_3$ zone axis, and the grain growth direction along [010] direction (green arrow). The blue boxed region is shown magnified as inset) (c) HRTEM lattice image from the blue box region in (b) confirming the growth direction of [010] and tangent direction of the bent film along [001] direction.*

As expected for a quasi-2D layer structure of $Sb_2S_3$ and locally indicated by the SAED pattern analysis, a highly [100] textured structure is obtained. To confirm this, we performed crystal orientation analysis using precession-assisted nano-beam 4D-STEM on the $Sb_2S_3$ thin film supported on the $SiO_x$ membrane (Fig. 3a), large area EBSD in the SEM (Fig. 3b for $Sb_2S_3$ and Fig. 3c for $ZnS/Sb_2S_3$) and polarization-dependent Raman spectroscopy (further detailed in Fig. 4 and text below) correlatively. In all cases, the inverse pole figures (IPF) along the z-axis (defined w.r.t. image coordinate, i e., out-of-plane direction) confirm the dominance of a [100] out-of-plane texture. The in-plane maps, e.g., IPF along image y-axis in Fig. 3a, clearly reveal that the long crystal grains extend along the [010] direction, consistent with the local SAED and HRTEM results above. We note that the noise pattern (i.e., small clusters of blue and red pixels among the dominance green pixels in the IPF z-axis) in Fig. 3a is due to strong bending of the film and failure of the orientation indexing routine at those locations, which is even more severe analysing the data without beam precession (cf. Fig. S4). Nevertheless, the few grains indexed to be close to [201] and [301] (orange and yellow grains as rendered in Fig. 3c IPF along z-axis) in the $ZnS/Sb_2S_3$ sample are carefully validated to be sample character. These grains can be mainly due to kinetics confined competition of grain growth, and dictated by the epitaxy of $Sb_2S_3$ on the ZnS layer, which was found to be randomly oriented nanopartricles from our earlier work [4]. These high-index oriented grains can create high carrier mobility channels in the out-of-plane direction, which is a desirable property for solar cell applications. [22]

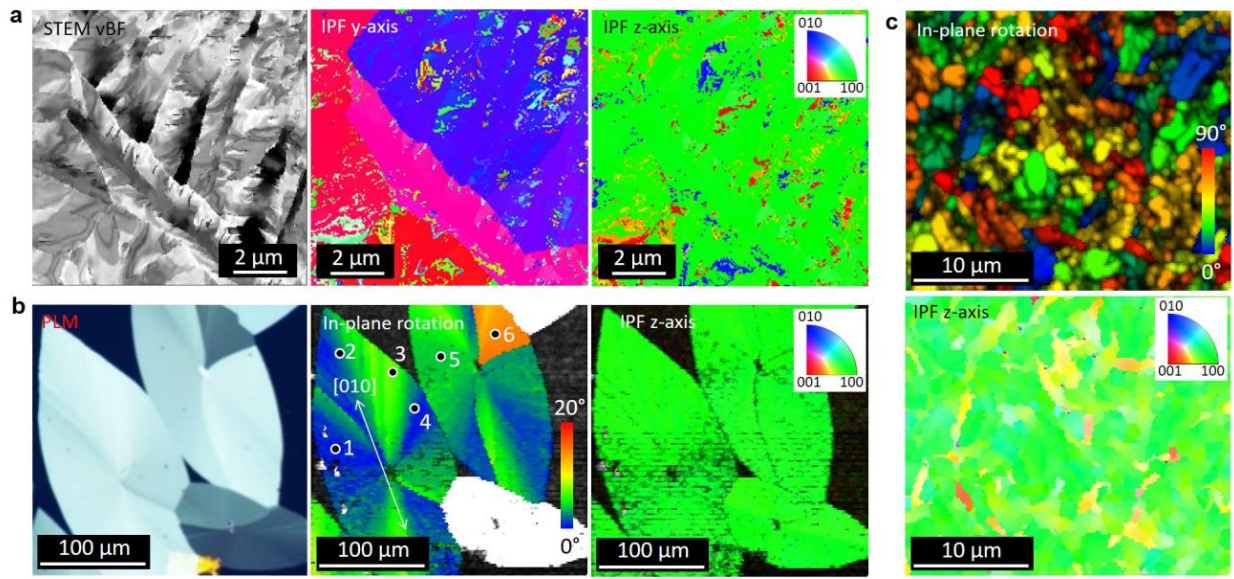

Figure 3 Crystal orientation analysis. (a) *Crystal orientation map with precession-assisted 4D-STEM dataset acquired in a thin window (blue box in the inset of Fig. 1e). (e) and (f) crystal orientation maps evaluated using EBSD of $Sb_2S_3$ crystallites on the Si/SiO$_x$ substrate as marked by the blue boxes in Fig. 1e and Fig. 1f, respectively. The in-plane rotation maps in (b) and (c) are presented a rainbow colour scheme to best visualize relative rotation angle of the [010] crystal direction of the crystals in the large leaf-like grains. The black dots (and numbers) in the map indicate where the Raman point measurements were carried out (shown in Fig. 4 and Fig. S3).*

The large leaf-like $Sb_2S_3$ crystallites grown on the $Si/SiO_x$ surface show also [100] textures. All grains are dominantly [010] oriented along the long axis of the "leaves", only exhibiting an in-plane rotation of up to 8°, which is also clearly reflected in the PLM (c.f. Fig. 3b). The smaller crystalline grains nucleated and grown on ZnS interlayer show also a dominant [100] out-of-plane texture. However, few grains appear to be [100] and [101] out-of-plane, which is likely due to the high density of nucleation and competitive crystal growth (further discussed later).

***Raman spectroscopy of the $Sb_2S_3$ thin films.*** We performed polarization-dependent Raman spectroscopy on the same sample as shown in Fig. 3b at spots with known in-plane rotation (indicated by the numbers in Fig. 3b), to correlatively probe the Raman response as function of the known crystal orientation. Figure 4a shows the Raman spectra of a "0° position" (spot 1 in Fig. 3b) with parallel polarization of incoming and detected light along the [010] in-plane direction (dark blue) and perpendicular to the [010] direction (middle and top, green lines). The most dominant $Sb_2S_3$ Raman modes are the *P1* peak (282 cm$^{-1}$) and the *P2* and *P3* peaks (301 cm$^{-1}$ and 312 cm$^{-1}$), marked by dashed lines. Depending on the orientation of the polarization, the intensity ratio of the *P1* to (*P2+P3)* peaks changes: for polarization along the [010] direction (long axis of the leaf-shape), the amplitude of the *P1* peak is higher than that of the (*P2/P3)* peaks, and vice versa for polarization perpendicular to the [010] axis. The same behavior is observed when the $Sb_2S_3$ sample is rotated by 90° instead of rotating the polarization of the laser and analyzer (dark green line in Fig. 4a).

At locations with slightly different in-plane rotations [5° (green/spot 3 and 5) and 12° (orange/spot 6), as indicated in the in-plane orientation map (Fig. 3b)], we observed small variations in the Raman spectra, specifically in the intensity ratio of the P2 to P3 peaks (see SI Fig. S3). As the in-plane rotation increases, the P2 peak becomes more prominent while the P1 peak decreases, eventually leading to the combined P2+P3 peak surpassing the intensity of the P1 peak.

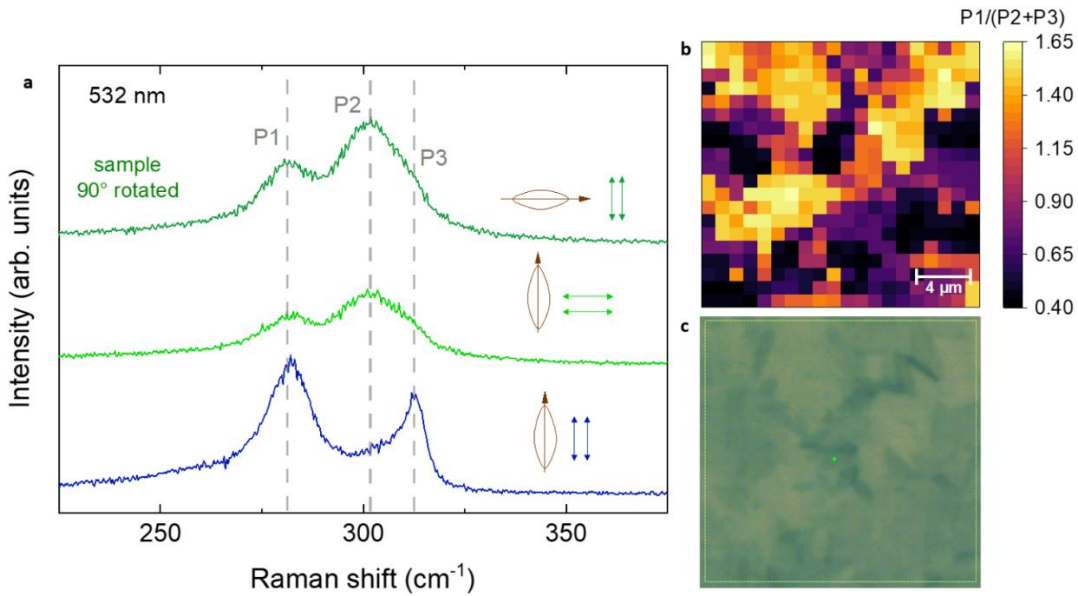

*Figure 4 Raman spectroscopy analysis of the $Sb_2S_3$ samples. (a) Raman spectra of $Sb_2S_3$ thin film with its [010] in-plane orientation being parallel (blue line), perpendicular (light green and dark green line) to the laser and analyzed polarization. (b) Raman map of the $ZnS/Sb_2S_3$ film showing the amplitude ratio of the P1 peak (~282 cm$^{-1}$) and the sum of P2 and P3 peaks (~301 cm$^{-1}$ and ~312 cm$^{-1}$). (c) Optical microscope image of the area measured in the Raman map in (b).*

Based on the amplitude ratio between the *P1* and (*P2+P3)* peaks indicating the crystalline orientation, we analyzed the ZnS/ $Sb_2S_3$ film. Due to the highly [100] out-of-plane texture and varying [010] in-plane orientation of the smaller crystalline grains on this sample (see Fig. 1f), strong polarization contrast was observed. We acquired a Raman map of 20x20 points over a field of view of 20x20 μm$^2$ (cf. Fig. 4b). Here, we plot the amplitude ratio of the *P1* to the (*P2+P3)* peaks. The polarization contrast in this map clearly reveals the crystal grains with different in-plane orientations, whereas their shape and size are in agreement with that revealed in Fig. 1f. The different directions can be effectively distinguished from polarized Raman spectra.

***In situ TEM observations and growth kinetics of $Sb_2S_3$.*** Having established the crystal grain morphology and orientation character, we now come back to examine the growth kinetics as observed in the *in situ* experiments. Figure 5 summarizes some key frames of the *in situ* ADF-STEM observations. The full videos and simultaneously acquired BF-, ABF- and HAADF-STEM images are provided as supplementary documents. Since the amorphous to crystalline phase transformation is irreversible, and the field-of-view, observation frame rate and resolution (sampling size) are dependent on each other, it is a matter of luck to capture the initial nucleation at high spatial and temporal resolution.

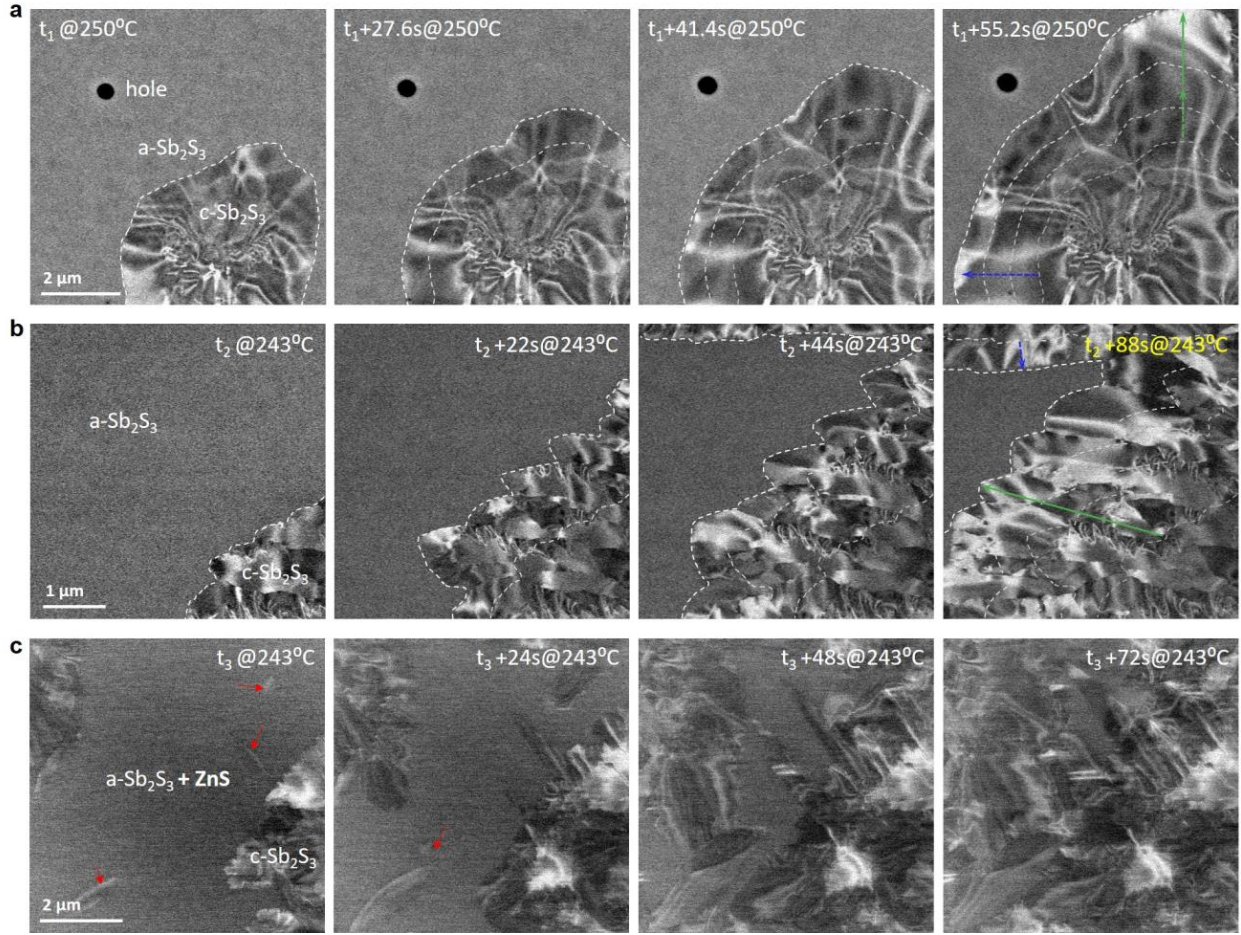

*Figure 5 in situ STEM observations of the nucleation of c-$Sb_2S_3$ and their growth from ALD a-$Sb_2S_3$ thin films (a, b) and from the ZnS/a-$Sb_2S_3$ film (c). (a) Snapshots extracted from experiment #1 (full movie in SI). Here, 2D isotropic growth of $Sb_2S_3$ at the initial nucleation site, as well as formation and expansion of a hole in the a-$Sb_2S_3$ film were observed. (b) Snapshots extracted from experiment #2 (full movie in SI). Growth of c-$Sb_2S_3$ far away from its initial nucleation site appeared as densely packed grains grown along common crystallographic direction (further analyzed in Fig. 3). (c) Snapshots extracted from experiment #3 (full movie in SI). At the same temperature as in exp. #2, higher density of nucleation sites was observed. No holes was found in the entire TEM grid after the phase transformation completed.*

Figure 5a highlights the growth front of a crystal grain from a nucleation site in the a-$Sb_2S_3$ thin film (without ZnS interlayer) at 250°C. It appears that the initial growth is isotropic in all directions with a growth rate of ~40 nm/s. Despite the strong bending contours, no clear indication of grain boundaries can be identified, suggesting the grain is still single crystalline. After about 30 seconds, the growth along the upward direction (in the image coordinate, indicated by the blue arrow) became faster at a growth rate of almost 100 nm/s. With the knowledge obtained from *ex situ* ACOM analysis, it is reasonable to assume the faster growth direction is along [010]. Assuming a constant growth rate hereafter and no coalescence, this would lead to oval shaped crystals with aspect ratio of about 2.5, which is in good agreement to the observed large "leaves" on the surface of the planar Si/$SiO_x$ substrate. Furthermore, a hole in the a-$Sb_2S_3$ region is observed. The diameter of the hole slightly grows with much slower kinetics as compared to the crystallization front.

Figure 5b showcases the growth front of crystal grains within the a-$Sb_2S_3$ thin film (without ZnS interlayer) at a slightly lower temperature of 243°C. The morphology of the grains mimics those of the characterization in Fig. 2d, thus it is reasonable to assume that the major growth direction (from right to left) is along [010]. Despite being only few degrees Celsius lower in temperature, this growth front extends at a much slower rate of about 35 nm/s as compared to the previous observations (up to 100 nm/s). From about $t_2$ + 40s another grain emerged from the top side of the frame with a growth rate of about 10 nm/s.

In the presence of an ultra-thin (1 nm) ZnS interlayer between the oxidic substrate and the antimony sulfide film, a much higher density of $Sb_2S_3$ nuclei are observed in the a-$Sb_2S_3$, as shown in the Fig. 5c (red arrows). Due to the higher density, grain growth is stopped by impinging neighboring grains before the typical leaf shaped growth characteristics evolve, resulting in a much finer grain structure as compared to the thin films without interlayer. Thus, the analysis of the growth kinetics reveal that the leaf shape morphology stems from the anisotropic growth kinetics of $Sb_2S_3$, here the growth is the fastest in [010] direction. In addition to this, the growth kinetics are highly temperature dependent leading to a significantly slower growth rate when reducing the crystallization temperature from 250 C to 243 C.

***Role of the ZnS interlayer on thin film degradation.*** During in situ annealing around 250°C, holes in the yet-to-transform a-$Sb_2S_3$ layer were observed in thin films without the ZnS interlayer (Fig. 5a and supplementary video 1), which is likely the very early stage of film degradation via solid state dewetting [23]. With longer annealing times at constant temperature around 250°C, these holes further expand. In the thin films with ZnS interlayer, no such holes were observed under identical conditions. It is interesting to note that the overall phase transformation kinetics with and without ZnS are rather similar. The higher density of nuclei with ZnS is in part countered by the fast anisotropic growth along [010] direction in the thin film without ZnS (c.f. Fig. 5a and 5b). Thus, we do not expect to kinetically suppress the thin film degradation/dewetting in the amorphous film, consequently the observed higher stability of the film on the ZnS interlayer can be attributed to a lowered interface energy between the a-$Sb_2S_3$ film and ZnS as compared to $SiO_x$. Those observations document that the role of the interlayer is twofold: i) ZnS provides heterogeneous nucleation sites leading to a finer and more isotropic microstructure after crystallization and ii) ZnS acts as an adhesion layer lowering the interface energy towards the substrate. However, it does not lower the phase transformation temperature nor the growth kinetics of $Sb_2S_3$ crystal growth.

## 3. Conclusion

This study provides a comprehensive understanding of the phase transformation, crystal growth, and degradation mechanisms of ALD-deposited **$Sb_2S_3$** thin films. Using *in situ* TEM we determined that the crystallization of amorphous **$Sb_2S_3$** films occurred at 243°C with anisotropic crystal growth, driven by its intrinsic structural anisotropy. Correlative microscopy and spectroscopy analyses confirmed the [100] out-of-plane texture and their orientation-dependent Raman signals. The introduction of an ultra-thin ZnS interfacial layer significantly enhanced nucleation density, leading to smaller, more uniform grains, while mitigating early-stage degradation. Conversely, ZnS-free films exhibited larger, leaf-shaped grains with holes formation and early stage degradation before phase transformation. Evaporation of the film took place at higher temperature of 350°C in vacuum, initiated along grain boundaries and proceeded preferentially along the [010] ribbon direction.

These findings highlight the critical roles of interfacial engineering and substrate properties in controlling crystallization dynamics and improving film stability. The combined use of *in situ* TEM, 4D-STEM, EBSD, and correlative optical spectroscopy provided a robust framework for unraveling complex growth phenomena, and bridging structural, chemical, and optical insights. By addressing the challenges of controlling nucleation, grain growth, and phase stability, this work establishes guidelines for tailoring

Sb$_2$S$_3$ thin films and offers valuable insights for their optimization in both optoelectronic and phase change applications.

**Materials and methods**

***Atomic layer deposition.*** ZnS and Sb$_2$S$_3$ were deposited using a homemade hot-wall atomic layer deposition (ALD) reactor onto Si/SiO$_x$ TEM grids with SiO$_x$ windows along the established protocols.[3, 4] The precursors used were diethylzinc (DEZ, 95%, abcr), tris(dimethylamido)antimony(III) (Sb(NMe$_2$)$_3$, 99.99%, Sigma-Aldrich), and H$_2$S (3% vol in N$_2$, Air liquide). Nitrogen was used as the carrier-gas, and the reaction temperatures were 150 and 120°C for the deposition of ZnS and Sb$_2$S$_3$, respectively. The precursors were kept at room temperature and the opening, exposure, and pumping times were 0.2, 15, and 15 s in all cases, except for the Sb precursor, which was opened for 1.5 s and kept at 40 °C due to the lower vapor pressure.

We note that ALD deposited a- Sb$_2$S$_3$ on both sides of the Si/SiO$_x$ TEM grid. In most of the SiOx windows, we observed overlapping Sb$_2$S$_3$ crystal films on both sides of the SiO$_x$ membrane (e.g. Fig. 1c). This overlap can be identified by the continuous bending contour (of the crystal on one side) which seemingly crossed the grain boundaries (of the crystals on the other side). Luckily, in the center and lower-right window of a TEM grid shown in Fig. 1e, only the a-Sb$_2$S$_3$ at one side of the SiOx membrane has transformed to Sb$_2$S$_3$ crystal (inset in Fig. 1e), while the other side remained amorphous. This structure made it more straightforward to analyze the crystal orientation of the crystals with 4D-STEM.

***Light microscopy.*** Polarized light microscopy (PLM) was carried out using a Nikon Eclipse LV100ND microscope using a linear polarizer L-Pol in condenser and an analyzer before the detector. During the experiment, the polarizer was rotated until the minimum reflection intensity from the known amorphous region was reach. This indicate the L-Pol and analyzer are perpendicular to each other. Afterwards, the samples were rotated until strongest optic contrast was revealed. Images were acquired using the Nikon R2i colour camera with auto white balance.

***In situ observations in the TEM.*** TEM experiments were performed either on a double Cs corrected Thermal Fisher Scientific (TFS) Titan Themis platform operated at 300 kV, or on a probe-corrected TFS Spectra 200 instrument. Since most imaging conditions are under-sampled nanoscale resolution (sampling), aberration correction was not critical. The HRTEM image shown in Fig. 2 was acquired with the image corrector (of the Titan Themis TEM) carefully tuned to deliver a point resolution of 1 Å.

In situ observations were made in STEM mode with probe convergence α of 21 mrad (on the spectra) and 15.7 mrad (on the Titan) and four imaging modalities, BF-, ABF-, ADF- and HAADF-STEM simultaneously recorded. The detector collection angles were set so that the BF detector receive up to α/2, ABF cover from α/2 to slightly > α, ADF from slightly > α to ~55 mrad, and HAADF receive from ~56 mrad to ~200 mrad. In such way, the strong diffraction contrast, bending contours can be best visualized in ABF- and ADF-STEM images even at high speed and noisy datasets, while the HAADF-STEM images can be used to identify grain boundaries, voids and holes.

A Gatan furnace heating holder system model 652 is used to perform the *in situ* annealing experiments. The temperature accuracy in steady state (e.g., during holding) was testified by quantitative evaluating dewetting phenomena of identical metal thin films compared to carefully calibrated MEMS chip-based heating holder, and the results show good agreement.[24] The Si/SiO$_x$ membrane windows grids were mounted so that the windows were aligned in the center of the heating furnace. The temperature was manually controlled by the controller knob. The homogeneous size of crystals found on the surface of the TEM grid after annealing experiments indicate no critical temperature gradient during the experiments.

***Crystal orientation mapping*** Precession-assisted nano-beam 4D-STEM experiments were performed using a TESCAN Tensor instrument operated at 120 kV. This instrument is equipped with a Dectris Quadro hybrid pixel detector and a precession unit capable to run up to 72 kHz. We applied focused probe of 2 mrad convergence angle and 250 pA probe current. Further acquisition parameters were: 1º precession angle, 1 ms dwell time 200x200 scanning points. Indexing and IPF are evaluated using TESCAN software with crystal structure of $Sb_2S_3$. Beam presession greatly improved the indexing success rate and generate orientation maps with higher conficence level (cf. supplementary Fig. S4 results with and without beam precession).

EBSD measurements were carried out on a Zeiss Gemini 560 equipped with an EDAX Clarity Super EBSD detector (AMETEK Inc., USA). The measurements were executed at 10kV, the probe current was set to 150-250 pA and a pixel size of 2 µm for the $Sb_2S_3$ and 0.2 µm for $Sb_2S_3$/ZnS. After data collection, the EBSD maps were refined via spherical indexing [25] in OIM Matrix (AMETEK) with a pre-calculated master pattern using EMsoft [26]. The master pattern from EMsoft greatly improve the indexing quality as compared to the native master pattern from OIM Matrix.

***Raman spectroscopy.*** Raman spectroscopy of the $Sb_2S_3$ and $Sb_2S_3$/ZnS samples was performed with a HORIBA LabRam HR Evolution spectrometer with a laser wavelength of $\lambda$ = 532.17 nm (2.33 eV), laser focus < 1 µm, and an 1800 lines/mm grating. To avoid heating effects, laser powers below 0.5 mW were used. For the polarization-dependent measurements, a polarizer and an analyzer were integrated into the beam path to control the polarization of the laser and the passing direction of the scattered light. For the Raman map of the $Sb_2S_3$/ZnS sample, a step size of 1 µm was used. All Raman spectra were calibrated by neon lines.

**References**

1. Lian, X.J., et al., *A review on the recent progress on photodetectors.* Journal of Materials Science, 2024. **59**(47): p. 21581-21604.
2. Kondrotas, R., C. Chen, and J. Tang, *Sb2S3 Solar Cells.* Joule, 2018. **2**(5): p. 857-878.
3. Büttner, P., et al., *Continuous, crystalline Sb2S3 ultrathin light absorber coatings in solar cells based on photonic concentric p-i-n heterojunctions.* Nano Energy, 2022. **103**: p. 107820.
4. Büttner, P., et al., *ZnS Ultrathin Interfacial Layers for Optimizing Carrier Management in Sb2S3-based Photovoltaics.* ACS Applied Materials & Interfaces, 2021. **13**(10): p. 11861-11868.
5. Kim, D.H., et al., *Highly reproducible planar Sb(2)S(3)-sensitized solar cells based on atomic layer deposition.* Nanoscale, 2014. **6**(23): p. 14549-54.
6. Chen, J.W., et al., *Recent Advances and Prospects of Solution-Processed Efficient Sb2S3 Solar Cells.* Advanced Functional Materials, 2024. **34**(18).
7. Zhao, Y.Q., et al., *A Review of Carrier Transport in High-Efficiency Sb2(S,Se)3 Solar Cells.* Solar Rrl, 2023. **7**(23).
8. Farhana, M.A., et al., *A review on the device efficiency limiting factors in Sb2S3-based solar cells and potential solutions to optimize the efficiency.* Optical and Quantum Electronics, 2023. **55**(8).
9. Farhana, M.A., A. Manjceevan, and J. Bandara, *Recent advances and new research trends in Sb2S3 thin film based solar cells.* Journal of Science-Advanced Materials and Devices, 2023. **8**(1).
10. Barthwal, S., R. Kumar, and S. Pathak, *Present Status and Future Perspective of Antimony Chalcogenide (Sb2X3) Photovoltaics.* Acs Applied Energy Materials, 2022. **5**(6): p. 6545-6585.
11. Li, T.T., et al., *Neuromorphic Photonics Based on Phase Change Materials.* Nanomaterials, 2023. **13**(11).

12. Tara, V., et al., *Non-Volatile Reconfigurable Transmissive Notch Filter Using Wide Bandgap Phase Change Material Antimony Sulfide.* Ieee Journal of Selected Topics in Quantum Electronics, 2024. **30**(4).
13. Chen, R., et al., *Non-volatile electrically programmable integrated photonics with a 5-bit operation.* Nature Communications, 2023. **14**(1).
14. Gutierrez, Y., et al., *Interlaboratory study on Sb(2)S(3) interplay between structure, dielectric function, and amorphous-to-crystalline phase change for photonics.* iScience, 2022. **25**(6): p. 104377.
15. Dong, W.L., et al., *Wide Bandgap Phase Change Material Tuned Visible Photonics.* Advanced Functional Materials, 2019. **29**(6).
16. Kepic, P., et al., *Pulsed laser deposition of Sb2S3 films for phase-change tunable nanophotonics.* New Journal of Physics, 2024. **26**(1): p. 013005.
17. Gao, K., et al., *Optimizing Wavelength for Enhanced Cycling Durability of Laser-Induced Phase Change in Sb2S3 Films: A Survey on Optical Transmission Phase Shift.* Advanced Optical Materials, 2024. **12**(13).
18. Guo, P.G., A.M. Sarangan, and I. Agha, *A Review of Germanium-Antimony-Telluride Phase Change Materials for Non-Volatile Memories and Optical Modulators.* Applied Sciences-Basel, 2019. **9**(3).
19. George, S.M., *Atomic Layer Deposition: An Overview.* Chemical Reviews, 2010. **110**(1): p. 111-131.
20. Xu, H.Y., et al., *Atomic layer deposition - approach to nanoscale hetero-interfacial engineering of chemical sensors electrodes: A review.* Sensors and Actuators B-Chemical, 2021. **331**.
21. Bayliss, P. and W. Nowacki, *Refinement of the crystal structure of stibnite, Sb2S3.* Zeitschrift für Kristallographie - Crystalline Materials, 1972. **135**: p. 308.
22. Jin, X., et al., *Controllable Solution-Phase Epitaxial Growth of Q1D Sb(2) (S,Se)(3) /CdS Heterojunction Solar Cell with 9.2% Efficiency.* Adv Mater, 2021. **33**(44): p. e2104346.
23. Thompson, C.V., *Solid-State Dewetting of Thin Films.* Annual Review of Materials Research, Vol 42, 2012. **42**: p. 399-434.
24. Niekiel, F., et al., *Local temperature measurement in TEM by parallel beam electron diffraction.* Ultramicroscopy, 2017. **176**: p. 161-169.
25. Lenthe, W.C., S. Singh, and M. De Graef, *A spherical harmonic transform approach to the indexing of electron back-scattered diffraction patterns.* Ultramicroscopy, 2019. **207**.
26. Singh, S., F. Ram, and M. De Graef, *EMsoft: open source software for electron diffraction/image simulations.* Microscopy and Microanalysis, 2017. **23**(S1): p. 212-213.

**Acknowledgement**

We acknowledge funding from Deutsche Forschungsgemeinschaft (DFG) via the research training school GRK 1896: "In-Situ Microscopy with Electrons, X-rays and Scanning Probes", and the Cluster of Excellence EXC 315 "Engineering of Advanced Materials" the electron microscopy instruments available at CENEM in Erlangen. M.K.S.B., V.M.K. and J.B. acknowledge funding by the Bavarian-Czech Academic Alliance (BTHA) in the project "BaCzALD -- Bavarian-Czech alliance for photoactive films by solution atomic layer deposition" (grant number BTHA-JC-2024-2). T.D. and J.M. acknowledge support by the Deutsche Forschungsgemeinschaft (DFG, German Research Foundation), project number 447264071 (INST 90/1183-1 FUGG).

**Conflict of interest**

The authors declare no conflict of interest.

**Author Contributions**

M.W., J.W., J.B., and E.S. conceptualized the study and designed the experiments. M.W. performed the *in situ* TEM experiments, formal analysis and investigation, validated the study, wrote the original draft, wrote, reviewed and edited the final draft. M.K.S.B. and V.M.K. performed the ALD sample preparation. M.D. performed the SEM-EBSD data acquisition and formal analysis. T.D. and J.M. performed the Raman measurement and formal analysis. P.L. coordinated and performed the precession-assisted 4D-STEM experiments and formal analysis. R.D.B. established instrumentation of precession-assisted 4D-STEM, reviewed and edited the original draft. J.M., J.B. and E.S. acquired funds, reviewed and edited the original draft. All authors approved the final draft.

**Data availability**

Data will be uploaded to public data repository and made available before final publication.

**Supplementary information**

Supplementary information are supplied as separate document.